\documentclass[preprint]{aastex}

\newcommand{\kcorrection}{$K$~correction}
\newcommand{\kcorrections}{{\kcorrection}s}
\newcommand{\nuobs}{\nu_o}
\newcommand{\nuemit}{\nu_e}
\newcommand{\lambdaobs}{\lambda_o}
\newcommand{\lambdaemit}{\lambda_e}

\newcommand{\latin}[1]{\textsl{#1}}
\providecommand{\ie}{\latin{i.e.}}
\providecommand{\eg}{\latin{e.g.}}

\begin{document}

\title{\scalebox{1.5}{The \kcorrection}}
\author{
  David W. Hogg\altaffilmark{1,2},
  Ivan K. Baldry\altaffilmark{3},
  Michael R. Blanton\altaffilmark{1},
  and
  Daniel J. Eisenstein\altaffilmark{4}
}
\altaffiltext{1}{
  Center for Cosmology and Particle Physics,
  Department of Physics,
  New York University
}
\altaffiltext{2}{
  \texttt{david.hogg@nyu.edu}
}
\altaffiltext{3}{
  Department of Physics and Astronomy,
  The Johns Hopkins University
}
\altaffiltext{4}{
  Steward Observatory,
  University of Arizona
}

\begin{abstract}
The \kcorrection\ ``corrects'' for the fact that sources observed at
different redshifts are, in general, compared with standards or each
other at different rest-frame wavelengths.  It is part of the relation
between the emitted- or rest-frame absolute magnitude of a source in
one broad photometric bandpass to the observed-frame apparent
magnitude of the same source in another broad bandpass.  This short
pedagogical paper provides definitions of and equations for the
\kcorrection.
\end{abstract}

\section{Introduction}

The expansion of the Universe provides astronomers with the benefit
that recession velocities can be translated into radial distances.  It
also presents the challenge that sources observed at different
redshifts are sampled, by any particular instrument, at different
rest-frame frequencies.  The transformations between observed and
rest-frame broad-band photometric measurements involve terms known as
``\kcorrections'' \citep*{humason56a, oke68a}.

Here we define the \kcorrection\ and give equations for its
calculation, with the goals of explanation, clarification, and
standardization of terminology.

In what follows, we consider a source observed at redshift $z$,
meaning that a photon observed to have frequency $\nuobs$ was emitted
by the source at frequency $\nuemit$ with
\begin{equation}
\nuemit = [1+z]\,\nuobs \;\;\;.
\end{equation}
The apparent flux of the source is imagined to be measured through a
finite observed-frame bandpass $R$ and the intrinsic luminosity is
imagined to be measured through a finite emitted-frame bandpass $Q$.
The \kcorrection\ is used in relating these two quantites.

Technically, the \kcorrection\ described here includes a slight
generalization from the original conception: The observed and
emitted-frame bandpasses are permitted to have arbitrarily different
shapes and positions in frequency space \citep[as they are in,
\eg,][]{kim96a}.  In addition, the equations below permit the different
bandpasses to be calibrated to different standard sources.

\section{Equations}

Consider a source observed to have apparent magnitude $m_R$ when
observed through photometric bandpass $R$, for which one wishes to
know its absolute magnitude $M_Q$ in emitted-frame bandpass $Q$.  The
\kcorrection\ $K_{QR}$ for this source is \emph{defined} by
\begin{equation}
\label{eq:definition}
m_R = M_Q + DM + K_{QR} \;\;\;,
\end{equation}
where $DM$ is the distance modulus, defined by
\begin{equation}
DM = 5\,\log_{10}\left[\frac{D_L}{10~\mathrm{pc}}\right] \;\;\;,
\end{equation}
where $D_L$ is the luminosity distance \citep[\eg,][]{hogg99cosm} and
$1~\mathrm{pc}= 3.086\times 10^{16}~\mathrm{m}$.

The apparent magnitude $m_R$ of the source is related to its spectral
density of flux $f_{\nu}(\nu)$ (energy per unit time per unit area per
unit frequency) by
\begin{equation}
m_R = -2.5\,\log_{10}\left[
  \frac{\displaystyle
          \int\frac{\mathrm{d}\nuobs}{\nuobs}\,f_{\nu}(\nuobs)\,R(\nuobs)}
       {\displaystyle
          \int\frac{\mathrm{d}\nuobs}{\nuobs}\,g^R_{\nu}(\nuobs)\,R(\nuobs)}
\right] \;\;\;,
\end{equation}
where the integrals are over the observed frequencies $\nuobs$;
$g^{R}_{\nu}(\nu)$ is the spectral density of flux for the
zero-magnitude or ``standard'' source, which, for Vega-relative
magnitudes, is Vega (or perhaps a weighted sum of a certain set of A0
stars), and, for AB magnitudes \citep{oke83a}, is a hypothetical
constant source with $g^\mathrm{AB}_{\nu}(\nu)=3631~\mathrm{Jy}$
(where $1~\mathrm{Jy}= 10^{-26}~\mathrm{W\,m^{-2}\,Hz^{-1}}=
10^{-23}~\mathrm{erg\,cm^{-2}\,s^{-1}\,Hz^{-1}}$) at all frequencies
$\nu$; and $R(\nu)$ describes the bandpass, as follows:

The value of $R(\nu)$ at each freqency $\nu$ is the mean contribution
of a photon of frequency $\nu$ to the output signal from the detector.
If the detector is a photon counter, like a CCD, then $R(\nu)$ is just
the probability that a photon of frequency $\nuobs$ gets counted.  If
the detector is a bolometer or calorimeter, then $R(\nu)$ is the
energy deposition $h\,\nu$ per photon times the fraction of photons of
energy $\nu$ that get absorbed into the detector.  If $R(\nu)$ has
been properly computed, there is no need to write different integrals
for photon counters and bolometers.  Note that there is an implicit
assumption here that detector nonlinearities have been corrected.

The absolute magnitude $M_Q$ is defined to be the apparent magnitude
that the source \emph{would} have if it were $10~\mathrm{pc}$ away, at
rest (\ie, not redshifted), and compact.  It is related to the
spectral density of the luminosity $L_{\nu}(\nu)$ (energy per unit
time per unit frequency) of the source by
\begin{equation}
M_Q = -2.5\,\log_{10}\left[
  \frac{\displaystyle
          \int\frac{\mathrm{d}\nuemit}{\nuemit}\,
              \frac{L_{\nu}(\nuemit)}{4\pi\,(10~\mathrm{pc})^2}\,Q(\nuemit)}
       {\displaystyle
          \int\frac{\mathrm{d}\nuemit}{\nuemit}\,g^Q_{\nu}(\nuemit)\,Q(\nuemit)}
\right] \;\;\;,
\end{equation}
where the integrals are over emitted (\ie, rest-frame) frequencies
$\nuemit$, $D_L$ is the luminosity distance, and $Q(\nu)$ is the
equivalent of $R(\nu)$ but for the bandpass $Q$.  As mentioned above,
this does not require $Q=R$, so this will lead to, technically, a
generalization of the \kcorrection.  In addition, since the $Q$ and
$R$ bands can be zero-pointed to different standard sources (\eg, if
$R$ is Vega-relative and $Q$ is AB), it is not necessary that
$g^Q_{\nu}=g^R_{\nu}$.

If the source is at redshift $z$, then its luminosity is related to
its flux by
\begin{equation}
\label{eq:luminosity}
L_{\nu}(\nuemit) = \frac{4\pi\,D_L^2}{1+z}\,f_{\nu}(\nuobs) \;\;\;,
\end{equation}
\begin{equation}
\nuemit = [1+z]\,\nuobs \;\;\;.
\end{equation}
The factor of $(1+z)$ in the luminosity expression
(\ref{eq:luminosity}) accounts for the fact that the flux and
luminosity are not bolometric but densities per unit freqency.  The
factor would appear in the numerator if the expression related flux
and luminosity densities per unit wavelength.

Equation (\ref{eq:definition}) holds if the \kcorrection\ $K_{QR}$ is
\begin{equation}
\label{eq:kcorrection}
K_{QR} = -2.5\,\log_{10}\left[[1+z]\,
  \frac{\displaystyle
          \int\frac{\mathrm{d}\nuobs}{\nuobs}\,f_{\nu}(\nuobs)\,R(\nuobs)\,
          \int\frac{\mathrm{d}\nuemit}{\nuemit}\,g^Q_{\nu}(\nuemit)\,Q(\nuemit)}
       {\displaystyle
          \int\frac{\mathrm{d}\nuobs}{\nuobs}\,g^R_{\nu}(\nuobs)\,R(\nuobs)\,
          \int\frac{\mathrm{d}\nuemit}{\nuemit}\,
            f_{\nu}\!\left(\frac{\nuemit}{1+z}\right)\,Q(\nuemit)}
\right] \;\;\;.
\end{equation}

Equation (\ref{eq:kcorrection}) can be taken to be an operational
definition, therefore, of the \kcorrection, from observations through
bandpass $R$ of a source whose absolute magnitude $M_Q$ through
bandpass $Q$ is desired.  Note that if the $R$ and $Q$ have different
zero-point definitions, the $g^R_{\nu}(\nuemit)$ in the numerator will
be a different function from the $g^Q_{\nu}(\nuobs)$ in the
denominator.

In equation (\ref{eq:kcorrection}), the \kcorrection\ was defined in
terms of the apparent flux $f_{\nu}(\nu)$ in the observed frame.  This
is the direct observable.  Most past discussions of the \kcorrection\
\citep[\eg,][]{oke68a, kim96a} write equations for the \kcorrection\ in
terms of either the flux or luminosity in the emitted frame.
Transformation from observed-frame flux $f_{\nu}(\nuobs)$ to
emitted-frame luminosity $L_{\nu}(\nuemit)$ gives
\begin{equation}
\label{eq:kcorrectionL}
K_{QR} = -2.5\,\log_{10}\left[[1+z]\,
  \frac{\displaystyle
          \int\frac{\mathrm{d}\nuobs}{\nuobs}\,L_{\nu}([1+z]\nuobs)\,R(\nuobs)\,
          \int\frac{\mathrm{d}\nuemit}{\nuemit}\,g^Q_{\nu}(\nuemit)\,Q(\nuemit)}
       {\displaystyle
          \int\frac{\mathrm{d}\nuobs}{\nuobs}\,g^R_{\nu}(\nuobs)\,R(\nuobs)\,
          \int\frac{\mathrm{d}\nuemit}{\nuemit}\,
            L_{\nu}(\nuemit)\,Q(\nuemit)}
\right] \;\;\;.
\end{equation}

In the above, all calculations were performed in frequency units.  In
wavelength units, the spectral density of flux $f_{\nu}(\nu)$ per unit
frequency is replaced with the spectral density of flux
$f_{\lambda}(\lambda)$ per unit wavelength using
\begin{equation}
\nu\,f_{\nu}(\nu) = \lambda\,f_{\lambda}(\lambda) \;\;\;,
\end{equation}
\begin{equation}
\lambda\,\nu = c \;\;\;,
\end{equation}
where $c$ is the speed of light.  The \kcorrection\ becomes
\begin{equation}
\label{eq:wavelength}
K_{QR} = -2.5\,\log_{10}\left[\frac{1}{[1+z]}\,
  \frac{\displaystyle
  \int\mathrm{d}\lambdaobs\,\lambdaobs\,f_{\lambda}(\lambdaobs)\,R(\lambdaobs)\,
    \int\mathrm{d}\lambdaemit\,\lambdaemit\,
    g^Q_{\lambda}(\lambdaemit)\,Q(\lambdaemit)}
       {\displaystyle
  \int\mathrm{d}\lambdaobs\,\lambdaobs\,g^R_{\lambda}(\lambdaobs)\,R(\lambdaobs)\,
    \int\mathrm{d}\lambdaemit\,\lambdaemit\,
    f_{\lambda}([1+z]\lambdaemit)\,Q(\lambdaemit)}
\right] \;\;\;,
\end{equation}
where, again, $R(\lambda)$ is defined to be the mean contribution to
the detector signal in the $R$ bandpass for a photon of wavelength
$\lambda$ and $Q(\lambda)$ is defined similarly.  Note that the
hypothetical standard source for the AB magnitude system, with
$g^\mathrm{AB}_{\nu}(\nu)$ constant, has
$g^\mathrm{AB}_{\lambda}(\lambda)$ not constant but rather
$g^\mathrm{AB}_{\lambda}(\lambda)=
c\,\lambda^{-2}\,g^\mathrm{AB}_{\nu}(\nu)$.

Again, transformation from observed-frame flux
$f_{\lambda}(\lambdaobs)$ to emitted-frame luminosity
$L_{\lambda}(\lambdaemit)$ gives
\begin{equation}
\label{eq:wavelengthL}
K_{QR} = -2.5\,\log_{10}\left[\frac{1}{[1+z]}\,
  \frac{\displaystyle
  \int\mathrm{d}\lambdaobs\,\lambdaobs\,L_{\lambda}\!\left(\frac{\lambdaobs}{1+z}\right)\,R(\lambdaobs)\,
    \int\mathrm{d}\lambdaemit\,\lambdaemit\,
    g^Q_{\lambda}(\lambdaemit)\,Q(\lambdaemit)}
       {\displaystyle
  \int\mathrm{d}\lambdaobs\,\lambdaobs\,g^R_{\lambda}(\lambdaobs)\,R(\lambdaobs)\,
    \int\mathrm{d}\lambdaemit\,\lambdaemit\,
    L_{\lambda}(\lambdaemit)\,Q(\lambdaemit)}
\right] \;\;\;.
\end{equation}

Equation (\ref{eq:wavelengthL}) becomes identical to the equation for
$K$ in \citet{oke68a} if it is assumed that $Q=R$, that
$g^Q_{\nu}=g^R_{\nu}$, that the variables $\lambda_0$,
$F(\lambda)$, and $S_i(\lambda)$ in \citet{oke68a} are set to
\begin{eqnarray}\displaystyle
\lambda_0 & = & \lambdaemit \;\;\;, \nonumber \\
F(\lambda) & = & L_{\lambda}(\lambda) \;\;\;, \nonumber \\
S_i(\lambda) & = & \lambda\,R(\lambda) \;\;\;,
\end{eqnarray}
and that the integrand $\lambda$ is used differently in each of the
two integrals.  Similar transformations make the equations here
consistent with those of \citet{kim96a}, although they distinguish
between the classical \kcorrection\ and one computed for photon
counting devices (an unnecessary distinction); their most similar
equation is that given for $K^\mathrm{counts}_{xy}$.

\section{Discussion}

To compute an accurate \kcorrection, one needs an accurate description
of the source flux density $f_{\nu}(\nu)$, the standard-source flux
densities $g^R_{\nu}(\nu)$ and $g^Q_{\nu}(\nu)$, and the bandpass
functions $R(\nu)$ and $Q(\nu)$.  In most real astronomical
situations, none of these is known to better than a few percent, often
much worse.  Sometimes, use of the AB system seems reassuring
(relative to, say, a Vega-relative system) because
$g^\mathrm{AB}_{\nu}(\nu)$ is known (\ie, defined), but this is a false
sense: In fact the standard stars have been put on the AB system to
the best available accuracy.  This involves absolute spectrophotometry
of at least some standard stars, but this absolute flux information is
rarely known to better than a few percent.  The expected deviations of
the magnitudes given to the standard stars from a true AB system are
equivalent to uncertainties in $g^\mathrm{AB}_{\nu}(\nu)$.

The classical \kcorrection\ has $Q(\nu)=R(\nu)$ and
$g^Q_{\nu}(\nu)=g^R_{\nu}(\nu)$.  This eliminates the integrals over
the standard-source flux density $g^R_{\nu}(\nu)$.  However, it
requires good knowledge of the source flux density $f_{\nu}(\nu)$ if
the redshift is significant.  Many modern surveys try to get
$R(\nu)\sim Q([1+z]\nu)$ so as to weaken dependence on $f_{\nu}(\nu)$,
which can be complicated or unknown.  This requires good knowledge of
the absolute flux densities of the standard sources if the redshift is
significant.  This kind of absolute calibration is often uncertain at
the few-percent level or worse.

Note that if equation (\ref{eq:definition}) is taken to be the
definition of the \kcorrection, then the statement by \citet{oke68a}
that the \kcorrection\ ``would disappear if intensity measurements of
redshifted galaxies were made with a detector whose spectral
acceptance band was shifted by $1+z$ at all wavelengths'' becomes
incorrect; the correct statement is that the \kcorrection\ would not
depend on the source's spectrum $f_{\nu}(\nu)$.

\acknowledgements It is a pleasure to thank Bev Oke for useful
discussions.  This research made use of the NASA Astrophysics Data
System, and was partially supported by funding from NASA and NSF.

\bibliographystyle{hacked_apj}
\bibliography{apj-jour,ccpp}

\begin{thebibliography}{5}
\expandafter\ifx\csname natexlab\endcsname\relax\def\natexlab#1{#1}\fi

\bibitem[{Hogg(1999)}]{hogg99cosm}
Hogg, D.~W. 1999, Distance measures in cosmology, astro-ph/9905116

\bibitem[{Humason {et~al}(1956)Humason, Mayall, \& Sandage}]{humason56a}
Humason, M.~L., Mayall, N.~U., \& Sandage, A.~R. 1956, Redshifts and magnitudes
  of extragalactic nebulae, \aj, 97--162

\bibitem[{Kim {et~al}(1996)Kim, Goobar, \& Perlmutter}]{kim96a}
Kim, A., Goobar, A., \& Perlmutter, S. 1996, A generalized {$K$} correction for
  type {Ia} supernovae:\ {C}omparing {$R$}-band photometry beyond $z=0.2$ with
  {$B$}, {$V$}, and {$R$}-band nearby photometry, \pasp, 190--201

\bibitem[{Oke \& Gunn(1983)}]{oke83a}
Oke, J.~B. \& Gunn, J.~E. 1983, Secondary standard stars for absolute
  spectrophotometry, \apj, 713--717

\bibitem[{{Oke} \& {Sandage}(1968)}]{oke68a}
{Oke}, J.~B. \& {Sandage}, A. 1968, Energy distributions, {$K$}~corrections,
  and the {S}tebbins--{W}hitford effect for giant elliptical galaxies, \apj,
  21--32

\end{thebibliography}

\end{document}